# Microwave-to-optical frequency conversion based on the Lamb shift

Sergey A. Rashkovskiy

*Ishlinsky Institute for Problems in Mechanics of the Russian Academy of Sciences, Vernadskogo Ave., 101/1, Moscow, 119526, Russia*

We consider the microwave-to-optical frequency conversion based on the Lamb shift, fine structure and forbidden transitions. The theory of such a conversion is developed and the efficiency of the microwave-to-optical frequency conversion is calculated. We show that the ratio of the peak power of the induced optical emission to the power of the incident microwave, for a hydrogen atom, can reach $10^6$.

Efficient conversion of signals between the microwave and the optical domain is a key feature required in classical and quantum communication networks [1,2]. For such a conversion process, several schemes have been investigated, including, cold atoms [3], spin ensembles coupled to superconducting circuits [4,5], and trapped ions [6,7]. The highest conversion efficiency so far was reached via electro-optomechanical coupling, where a high-quality mechanical membrane [8,9] or a piezoelectric photonic crystal [10] provide the link between an electronic LC circuit and laser light. Efficient direct electro-optic modulation [11-15] is considered as an alternative approach.

The Lamb-Retherford experiments [16-21] suggest a direct method of the microwave-to-optical frequency conversion using the forbidden transitions and the Lamb shift.

Indeed, the Lamb-Rutherford experiments [16-21] are based on microwave stimulation of excited hydrogen atoms that are in a metastable $2s_{1/2}$ state, into which they were transferred under the action of an external excitation source (electron beam). The electric dipole spontaneous transition of the hydrogen atom from the excited $2s_{1/2}$ state to the ground state $1s_{1/2}$ is forbidden by the selection rule $\Delta l = \pm 1$. The allowed quadrupole spontaneous transition $2s_{1/2} - 1s_{1/2}$ is a very slow process, which ensures the lifetime of the excited $2s_{1/2}$ state of the order of 1/7 s; this is 8 orders of magnitude longer than the lifetime of the excited states $2p_{1/2}$ and $2p_{3/2}$, which is of the order of $1.6 \times 10^{-9}$ s.

According to the Schrodinger equation, the excited states $2s_{1/2}$, $2p_{1/2}$ and $2p_{3/2}$ are degenerate. Allowance for relativistic effects (the Dirac equation) removes degeneracies for the $2p_{3/2}$ level (the fine structure of the spectral lines), however, does not remove the degeneracy of the $2s_{1/2} - 2p_{1/2}$ levels [22]. The theory shows [22] that the degeneracy of the $2s_{1/2} - 2p_{1/2}$ levels is removed by taking into account the radiative corrections (Lamb shift). The frequency shift associated with the fine structure of the levels $2s_{1/2} - 2p_{3/2}$ is 10949 MHz, while the Lamb shift is (1057.77±0.10) MHz.

When irradiating hydrogen atoms that are in the long-lived excited state $2s_{1/2}$, with microwave radiation at a frequency of ~10949 MHz or at a frequency of ~1057.77 MHz, Lamb and Retherford induced transitions, respectively, $2s_{1/2} - 2p_{3/2}$ and $2s_{1/2} - 2p_{1/2}$, which transferred the atoms to the short-lived states $2p_{3/2}$ and $2p_{1/2}$. Due to the spontaneous transition, the $2p_{3/2}$ and $2p_{1/2}$ states were rapidly relaxed to the ground state $1s_{1/2}$; these transitions were accompanied by UV radiation with a wavelength of ~122 nm. Thus, in experiments [16-21], in fact, a direct conversion of microwave radiation to UV radiation was carried out using forbidden transitions and degenerate states of the atom.

The analysis of experiments [16-21] allows us to purposefully organize such a microwave-to-optical frequency conversion. The theory of experiments [16-21] was developed in [23]. In what follows we shall use the results of [23] and also of [24-26].

Let us consider a device that operates as follows. There is a container containing an active medium - atomic hydrogen. An external energy source pumps the active medium: hydrogen atoms are transferred from the ground (pure) state $1s_{1/2}$ into a metastable mixed state with two excited modes $1s_{1/2}$ and $2s_{1/2}$, whose lifetime is about of 1/7 s. The unstable modes $2p_{1/2}$ and $2p_{3/2}$ excited at the same time are rapidly decayed (in a time of the order of $10^{-9}$ s). Pumping can be carried out, for example, as in the Lamb-Retherford experiments, using an electron beam, and also an electric discharge, X-ray or ultraviolet radiation. After the required pump level is reached, the pump source is turned off and the microwave source is turned on at a frequency of 10949 MHz to induce "transition" $2s_{1/2} \rightarrow 2p_{3/2}$ or at a frequency of 1057.9 MHz to induce "transition" $2s_{1/2} \rightarrow 2p_{1/2}$. As a result, under the action of microwave radiation, the forced flow of the electric charge of the electron wave occurs from the mode $2s_{1/2}$ into the corresponding mode $2p$. In both cases, the resulting mixed excited state $1s_{1/2} - 2p$ decays rapidly due to spontaneous emission at a wavelength of ~122 nm, which differs for the two cases under consideration only in the magnitude of the corrections associated with the fine structure.

The optical radiation induced by microwave radiation will be forced: by changing the intensity of microwave radiation, one can control the intensity of optical radiation. In particular, turning off the source of microwave radiation one can reduce the intensity of optical radiation to practically zero (taking into account the long lifetime of the metastable mixed state $1s_{1/2} - 2s_{1/2}$, spontaneous emission at the $1s_{1/2} \rightarrow 2s_{1/2}$ "transition" frequency will be very weak).

We note that this process looks like a three-level laser, however, as it is easy to see, the principle of operation of the device under consideration differs fundamentally from the principle of laser operation.

Using the results [23-26], we calculate the intensity of the stimulated emission of a hydrogen atom which are in a mixed metastable state with an excited $2s_{1/2}$ mode, depending on the intensity of the microwave radiation that induces the $2s_{1/2} \to 2p_{3/2}$ transition.

According to classical electrodynamics [27], the intensity of the electric dipole radiation $I = \frac{2}{3c^3} \overline{\ddot{\mathbf{d}}^2}$, where $\mathbf{d}(t) = -e \int \mathbf{r}|\psi|^2 d\mathbf{r}$ is the electric dipole moment of the electron wave in the hydrogen atom; $\psi$ is the wave function of an electron wave for a hydrogen atom (solution of the Schrodinger equation).

For a three-level atom with three excited modes 1, 2, and 3, using the results of [23, 24], one obtains

$$I = \lambda \hbar \omega_{31} \frac{|b_{32}|^2}{2\gamma_{31}} (\rho_{22} - \rho_{33})$$

where mode 1 is the mode $1s_{1/2}$; mode 2 is the mode $2s_{1/2}$; mode 3 is the mode $2p_{3/2}$; $\omega_{nk} = \omega_n - \omega_k$; $\mathbf{d}_{nk} = \mathbf{d}_{kn}^* = -e \int \mathbf{r} u_n^*(\mathbf{r}) u_k(\mathbf{r}) d\mathbf{r}$

$$\gamma_{nk} = \frac{2\omega_{nk}^3}{3\hbar c^3} |\mathbf{d}_{nk}|^2 \tag{1}$$

$$b_{nk} = b_{kn}^* = \frac{1}{\hbar}(\mathbf{d}_{nk} \mathbf{E}_0) \tag{2}$$

$$\rho_{22} = \rho_{22}(0) \exp\left\{-\frac{|b_{32}|^2}{2\gamma_{31}} \lambda t\right\} \tag{3}$$

$$\lambda = \frac{\gamma_{31}^2}{\gamma_{31}^2 + (\omega_{32}+\omega)^2} + \frac{\gamma_{31}^2}{\gamma_{31}^2 + (\omega_{32}-\omega)^2}$$

is the damping decrement of the stimulated emission; $\mathbf{E}_0$ and $\omega$ are the amplitude and frequency of microwave radiation; $\rho_{nn}$ is the degree of excitation of the mode $n$, such that $-e\rho_{nn}$ is equal to the electric charge of the electron wave contained in the mode $n$ [23-26]; $\omega_n$ and $u_n(\mathbf{r})$ are the eigenfrequency and eigenfunction of the Schrödinger equation for the mode $n$; $\rho_{22}(0)$ is the initial degree of excitation of the $2s_{1/2}$ mode.

For the weakly excited mode $2s_{1/2}$, we have $\rho_{33}/\rho_{22} \ll 1$ [23]; as a result, we obtain

$$I \approx \lambda \hbar \omega_{31} \frac{|b_{32}|^2}{2\gamma_{31}} \rho_{22} \tag{4}$$

We introduce

$$(\mathbf{d}_{32} \mathbf{E}_0) = |\mathbf{d}_{32}||\mathbf{E}_0| \cos \theta \tag{5}$$

where $\theta$ is the angle between the vectors $\mathbf{d}_{nk}$ and $\mathbf{E}_0$.

Then, using (1), (2) and (5), one obtains

$$I \approx \lambda \frac{3c^3}{4\omega_{31}^2} \frac{|\mathbf{d}_{32}|^2}{|\mathbf{d}_{31}|^2} |\mathbf{E}_0|^2 \cos^2 \theta \, \rho_{22}$$

The energy flux density of microwave radiation $S_{mw} = \frac{c}{8\pi}|\mathbf{E}_0|^2$, because of this

$$I \approx \lambda \frac{6\pi c^2}{\omega_{31}^2} \frac{|\mathbf{d}_{32}|^2}{|\mathbf{d}_{31}|^2} \cos^2\theta \, \rho_{22} S_{mw} \tag{6}$$

Relation (6) describes the intensity of stimulated emission of one atom, which can be characterized by the effective cross section of stimulated emission $\sigma = I/S_{mw}$. Thus, microwave-to-optical frequency conversion can be realized using even a single atom.

Let us consider the case when the active medium contains a set of atoms with a random orientation $\theta$. In this case, averaging (6) over $\theta$, we obtain the total intensity of stimulated emission

$$\bar{I}_\Sigma \approx \lambda N \frac{6\pi c^2}{\omega_{31}^2} \frac{|\mathbf{d}_{32}|^2}{|\mathbf{d}_{31}|^2} \overline{\rho_{22} \cos^2\theta} \, S_{mw}$$

where $N$ is the number of atoms;

$$\overline{\rho_{22} \cos^2\theta} = \frac{1}{2}\int_0^\pi \rho_{22}(\theta) \cos^2\theta \sin\theta \, d\theta$$

Taking into account equations (3) and (5), one obtains

$$\overline{\rho_{22} \cos^2\theta} = \rho_{22}(0) \int_0^1 x^2 \exp\{-\beta x^2\} dx$$

where $\beta = \frac{3|\mathbf{E}_0|^2}{32\pi^3\hbar}\lambda_{31}^2 \frac{|\mathbf{d}_{32}|^2}{|\mathbf{d}_{31}|^2}\lambda t$; $\lambda_{31} = 2\pi c/\omega_{31}$ is the wavelength of radiation at frequency $\omega_{31}$.

As a result, we obtain

$$\bar{I}_\Sigma \approx \lambda N \frac{3}{2\pi} \lambda_{31}^2 \frac{|\mathbf{d}_{32}|^2}{|\mathbf{d}_{31}|^2} \rho_{22}(0) f(\beta) S_{mw}$$

where

$$f(\beta) = \int_0^1 x^2 \exp\{-\beta x^2\} dx \tag{7}$$

The dependence of the integral (7) on the parameter $\beta$ is shown in Fig. 1.

From the analysis of the function $f(\beta)$ it follows that the degeneracy of the stimulated emission occurs non-exponentially with time. In particular, for $\beta < 4$, the integral (7) is well approximated by the function $f(\beta) = \frac{1}{3}\exp(-\beta/2)$, while for $\beta \gg 1$ there is an asymptotic behavior:

$$f(\beta) = \int_0^1 x^2 \exp\{-\beta x^2\} dx = \beta^{-3/2} \int_0^{\sqrt{\beta}} y^2 \exp\{-y^2\} dy \approx \beta^{-3/2} \int_0^\infty y^2 \exp\{-y^2\} dy = \frac{\sqrt{\pi}}{4}\beta^{-3/2}$$

This asymptotic describes the integral $f(\beta)$ well at $\beta \geq 4$. When the parameter $\beta$ changes from zero to $\beta \approx 6$, the value of the integral (7) drops approximately 10 times. Proceeding from this, we can estimate the characteristic time of degeneracy of the radiation

$$\tau = 2 \times 10^3 \frac{\hbar}{\lambda |\mathbf{E}_0|^2 \lambda_{31}^3} \frac{|\mathbf{d}_{31}|^2}{|\mathbf{d}_{32}|^2}$$

which is inversely proportional to the intensity of the incident microwave radiation.

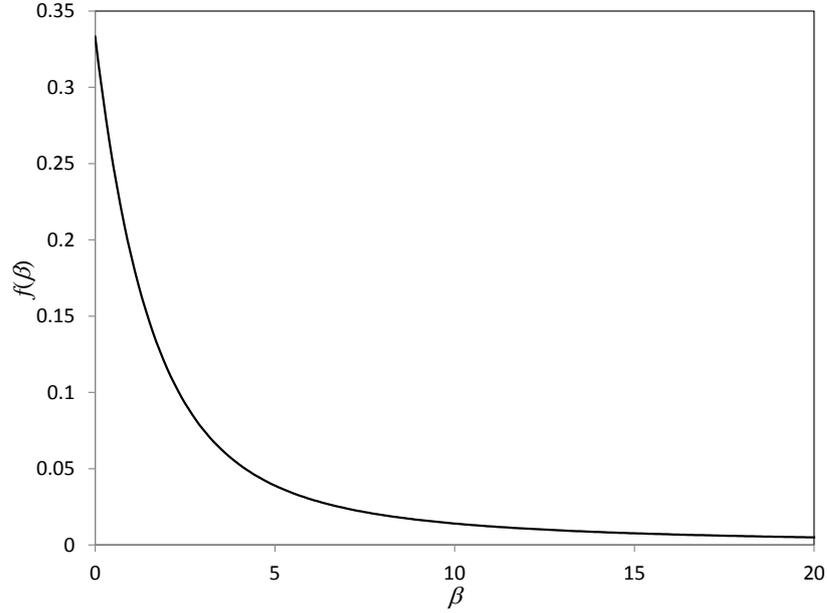

Fig. 1. Dependence of the integral (7) on the parameter $\beta$.

The radiation characteristic in this case is the effective cross section of the stimulated emission $\sigma_\Sigma = \bar{I}_\Sigma / S_{mw}$. As a result, we obtain

$$\sigma_\Sigma = \lambda N \frac{3}{2\pi} \lambda_{31}^2 \frac{|\mathbf{d}_{32}|^2}{|\mathbf{d}_{31}|^2} \rho_{22}(0) f(\beta)$$

The maximum effective cross section $\sigma_{\max}$ of such a source is achieved at a resonant frequency $\omega_{\max}$, at which $\lambda_{\max} \approx 1$ [23].

Thus

$$\sigma_{\max} \approx N \frac{3}{2\pi} \lambda_{31}^2 \frac{|\mathbf{d}_{32}|^2}{|\mathbf{d}_{31}|^2} \rho_{22}(0) f(\beta)$$

Let the active medium be a cylindrical vessel of length $L$ and cross-sectional area $F$ filled with hydrogen of density $\rho_H$. After activation (pumping), microwave radiation is applied into the cross section of this cylinder. The intensity of the microwave radiation passing through the cylinder is $FS_{mw}$. The ratio of the maximum intensity (power) of the stimulated emission to the intensity of the primary microwave radiation passing through the cylinder is $\eta_{\max} = \sigma_{\max}/F$, or

$$\eta_{\max} \approx N \frac{3}{2\pi} \frac{\lambda_{31}^2}{F} \frac{|\mathbf{d}_{32}|^2}{|\mathbf{d}_{31}|^2} \rho_{22}(0) f(\beta)$$

Taking into account that $N = m_H/\mu_H = \rho_H F L/\mu_H$, where $m_H$ $m_H$ is the mass of hydrogen in the vessel, $\mu_H$ is the atomic mass of hydrogen, we obtain

$$\eta_{\max} \approx \frac{3}{2\pi} N_{31} \frac{|\mathbf{d}_{32}|^2}{|\mathbf{d}_{31}|^2} \rho_{22}(0) f(\beta)$$

where $N_{31} = \rho_H L \lambda_{31}^2 / \mu_H$.

To estimate, we take, for example, $L = 10$ cm, $\rho_H = 0.9 \times 10^{-4}$ g/cm$^3$; $\lambda_{31} = 122$ nm; $\mu_H = 1.67 \cdot 10^{-24}$ g. As a result, we obtain $N_{31} = 0.8 \times 10^{11}$.

Then

$$\eta_{\max} \approx 4 \times 10^{10} \frac{|\mathbf{d}_{32}|^2}{|\mathbf{d}_{31}|^2} \rho_{22}(0) f(\beta)$$

Thus, even at a small initial degree of excitation of the $2s_{1/2}$ mode ($\rho_{22}(0) \ll 1$), very high stimulated emission efficiency can be obtained. For example, if $\rho_{22}(0) = 10^{-4}$ is provided, then, taking $\frac{|\mathbf{d}_{32}|^2}{|\mathbf{d}_{31}|^2} \sim 1$, we obtain $\eta_{\max} \sim 4 \times 10^6 f(\beta)$; i.e. the peak intensity of the stimulated emission is a million times greater than the intensity of the primary microwave radiation that causes it.

The expressions obtained are valid for the case when a fine structure of the $2p$-$1s$ spectral line of the hydrogen atom is used for microwave-to-optical frequency conversion. In order to describe the microwave-to-optical frequency conversion based on the Lamb shift $2s_{1/2} - 2p_{1/2}$, it suffices to simply interchange the indices 2 and 3 in the corresponding expressions and consider that the mode 2 is the $2p_{1/2}$ mode while the mode 3 is the $2s_{1/2}$ mode [23]. It is evident that in both cases under consideration, the efficiency $\eta_{\max}$ of the microwave-to-optical frequency conversion is the same.

As a conclusion, we note that microwave-to-optical frequency conversion can be performed not only on the Lamb shift or on the fine structure of the $2p_{1/2}$ - $1s_{1/2}$ spectral line, but also on any degenerate spectral line containing "transitions" with a long relaxation time. Such a source of radiation can be especially effective on "forbidden transitions", when an electron wave can be forced to "accumulate" in eigenmode, from which the electric dipole transition to the ground state (via any channels) is forbidden by selection rules. This will allow "accumulating light" in the metastable states of atoms and "store" it for a long time, and when necessary, quickly "extract" it under an external action of a weak microwave.

**Acknowledgements.** This work was done on the theme of the State Task No. AAAA-A17-117021310385-6.